\documentclass[11pt,twoside]{article}

\usepackage{asp2006}
\usepackage{epsf}
\usepackage{psfig}
\usepackage{lscape}

\newcommand{\be}{\begin{equation}}
\newcommand{\ee}{\end{equation}}
\newcommand{\rn}[1]{(\ref{#1})}
\newcommand{\ms}{{\rm m\, s}^{-1}}
\newcommand{\epsfb}[1]{\epsfxsize=4.8in \centerline{\epsffile{#1}}}
\newcommand{\epsf}[1]{\epsfxsize=5in \centerline{\epsffile{#1}}}

\begin{document}
\title{Inflated planets and their low-mass companions}   
\author{Rosemary A.\ Mardling}   
\affil{School of Mathematical Sciences, Monash University, Victoria, 3800, Australia}    

\begin{abstract} 
Various mechanisms have been proposed to explain the inflated size of HD~209458b
after it became clear that it has no companions capable of producing a stellar reflex
velocity greater than around $5\,\ms$. 
Had there been such a companion, the hypothesis 
that it forces the eccentricity of the inflated planet thereby tidally heating it
may have been readily accepted.
Here we summarize a paper by the author 
which shows that companion planets with masses as low
as a fraction of an Earth mass are capable of sustaining a non-zero eccentricity in
the observed planet for at least the age of the system. While such companions produce
stellar reflex velocities which are fractions of a meter per second and hence are
below the stellar jitter limit, they are consistent with recent theoretical work which
suggests that the planet migration process often produces low-mass companions
to short-period giants.

\end{abstract}

\section{Introduction}
Of the fourteen transiting extrasolar  
planetary systems for which radii have been measured, 
at least three appear to be
considerably larger than theoretical estimates suggest.
It has been proposed by \citet{b4} that
undetected companions acting to excite the orbital eccentricity
are responsible for these oversized planets, 
as they find new equilibrium radii in response to being tidally heated.
In the case of HD 209458, this hypothesis has been rejected by
some authors because
there is no sign of such a companion at the 5 ms$^{-1}$ level,
and because it is difficult to say conclusively that the eccentricity is non-zero.
Transit timing analysis as well
as a direct transit search have further constrained the existence of very short-period
companions, especially in resonant orbits.
Whether or not a companion is responsible for the large
radius of HD 209458b, almost certainly some short-period systems
have companions which force their eccentricities to
nonzero values. This paper is a summary of \citet{m1} which is 
dedicated to quantifying this effect.

The eccentricity of a short-period planet will
only be excited as long as its (non-resonant) companion's eccentricity is non-zero.
In fact, \citet{m1} shows that the latter decays on a timescale which
depends on the structure of the {\it interior} planet,
a timescale which is often shorter than the lifetime of the system.
{\it This includes Earth-mass planets in the habitable zones of some stars}.
On the other hand, there exist parameter combinations ($Q$-value of the innermost planet,
companion mass and semimajor axis) for which
significant eccentricity in the short-period planet can be 
sustained for at least the age of the system,
and these include systems with companion masses
as low as a fraction of an Earth mass.

\section{Inflated planets}

The three largest planets discovered to date are HD 209458b \citep{c2,h1}, 
HAT P1b  \citep{b1} and WASP-1b \citep{c1}. Table~\ref{planetdata}
\begin{table}[!ht]
\label{planetdata}
\caption{Inflated planet data}
\smallskip
\begin{center}
{\small
\begin{tabular}{lcccccc}
\tableline
\noalign{\smallskip}
& $a_p$ (AU) & $m_p\,\,(M_J)$ & $R_p\,\,(R_J)$ & $e_p$ & $\tau_{circ}$ (Gyr) & age (Gyr)\\
\noalign{\smallskip}
\tableline
\noalign{\smallskip}
HD 209458b & 0.045 & 0.64 & 1.32 & 0.014 & 0.14 & 5.5 \\
HAT P1b       & 0.055 & 0.53 & 1.36 & 0.09 & 0.33 & 3.6 \\
WASP 1b     & 0.038  & 0.79 & 1.44 & - & 0.03 & 2.0\\
\noalign{\smallskip}
\tableline
\end{tabular}
}
\end{center}
\end{table}
lists the measured attributes of these planets together
with the estimated age and circularization timescales, the latter assuming
a $Q$-value $Q_p=3\times 10^5$. It is clear from this data that if the eccentricities
are real, some mechanism operates to maintain them since $\tau_{circ}\ll{\rm age}$
in each case.

\citet{b3} and \cite{b2} calculated the rate at which energy needs to be
deposited into a gas giant with a specified mass and age
(and with and without a core) to maintain the observed radii of these planets.
However, they assumed that the energy is distributed uniformly throughout the
planet's envelope and dissipated locally; if the energy is instead transported 
and dissipated close (but not too close) to the surface, a smaller deposition
rate may be adequate. Consequently, \citet{b3} found that an eccentricity larger than that
observed for HD 209458b is necessary to maintain the planet's radius, at least
for reasonable $Q$-values.

Here we assume $Q_p=3\times 10^5$, a value consistent with the analysis of \citet{w3}
for all short-period gas giants discovered at that time, and assume that the estimated
eccentricities of the planets in Table~\ref{planetdata} are accurate.
We will demonstrate that it is possible for extremely low-mass planets to
maintain such eccentricities for at least the lifetime of the system.

\section{Long-term eccentricity evolution}

\begin{figure}[!h]
\epsfb{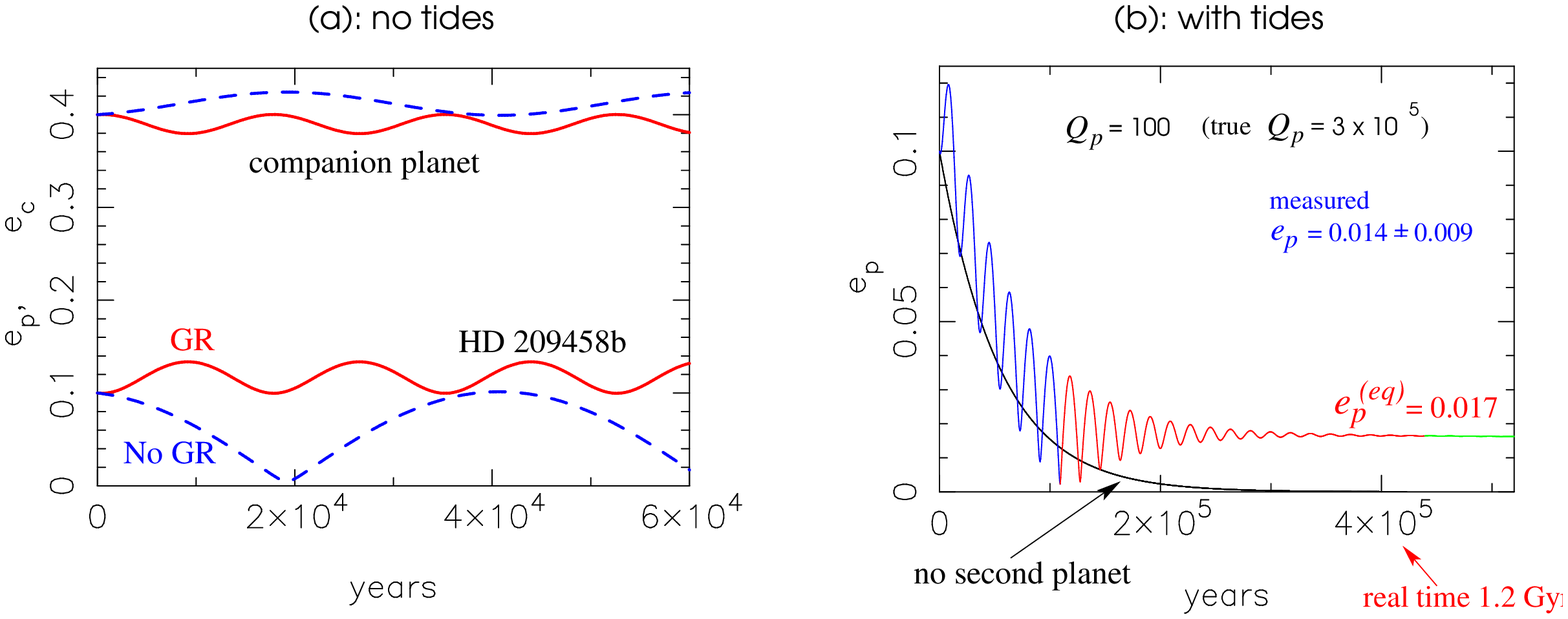}
 \caption{Secular evolution (a) without and (b) with tidal damping.}
 \label{secular}
 \end{figure}

Figure~\ref{secular}(a)
shows three-body secular evolution for an HD 209458b-like system with a 
companion planet of mass $0.1 M_J$ at 0.4 AU. The eccentricity of both planets varies
periodically with the period reducing when the post-Newtonian potential of
the star is included.
Figure~\ref{secular}(b)
 shows the evolution of the eccentricity of HD 209458b with the same companion but with tidal damping
 included. Here an artificially low value of $Q_p=100$
 was chosen to clearly illustrate the behavior. 
 This time the eccentricity appears to evolve to a fixed value on a timescale of around
 $1.4\times 10^5$ yr,
 which for $Q_p=3\times 10^5$ corresponds to 0.42 Gyr 
 ($3\tau_{circ}$; see \citet{m1} for details).
 The ``equilibrium'' or ``fixed-point'' eccentricity is given by
\be
e_p^{(eq)}=\frac{(5/4)(a_p/a_c)\,e_c\,\varepsilon_c^{-2}}
{\left|1-\sqrt{a_p/a_c}(m_p/m_c)\varepsilon_c^{-1}+\gamma\varepsilon_c^3\right|},
\label{equilGR}
\ee
where $a_p$ and $m_p$ are the semimajor axis and mass of the observed planet,
$a_c$ and $m_c$ are those of the hypothetical companion, $e_c$ is its eccentricity,
$\varepsilon_c=\sqrt{1-e_c^2}$ and $\gamma=4(n_p a_p/c)^2(m_*/m_c)(a_c/a_p)^3$.
Here $n_p$ is the observed planet's mean motion, $m_*$ is the star's mass and
$c$ is the speed of light. Capture by the fixed point is also accompanied by a transition
from circulation to libration of the angle between the lines of apsides of the orbits.

\begin{figure}[!h]
\epsf{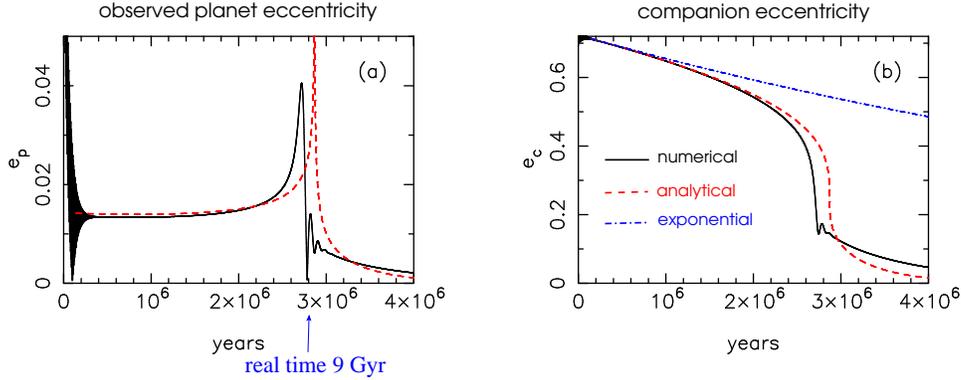}
 \caption{Pathological behavior of systems with very low-mass companions
 (see text for discussion).}
 \label{earth}
 \end{figure}

While the system evolves to a (psuedo)-fixed point on the timescale $3\tau_{circ}$
(with very little change in $e_c$), in the long run
the true equilibrium or fixed-point state of the system is two circular orbits.
This occurs on a timescale which is given approximately by
\be
\tau_c=(16/25)(m_c/m_p)(a_c/a_p)^{5/2}F(e_c^*)^{-1}\tau_{circ},
\label{tauc}
\ee
where $e_c^*$ is the value of $e_c$ when the system first finds $e_p^{(eq)}$, and
$F(e_c)=\varepsilon_c^{3}(1-\alpha\varepsilon_c^{-1}
+\gamma\,\varepsilon_c^3)^{2}$ with $\alpha=\sqrt{a_p/a_c}(m_p/m_c)$.
In fact, in general \rn{tauc} tends to underestimate the true circularization time
of the system, and a much more precise estimate is available \citep{m1}.
The analysis leading to this also reveals that a different
type of solution exists for very low-mass companions. Such a solution does not involve
slow exponential decay of the eccentricities, but rather a slow
{\it increase} of the observed planet's eccentricity to a maximum value
before rapid decay to zero.
Figure~\ref{earth}
demonstrates this behavior for a system with a companion mass of $4 M_\oplus$,
$a_c=0.3$ AU and $e_c(0)=0.7$.
Such solutions have parameters which make $e_p^{(eq)}$ (equation~\rn{equilGR})
near singular. Also shown is an analytical solution \citep{m1} and the
approximate exponential solution with timescale \rn{tauc}. Clearly the latter now
{\it overestimates} the timescale, however, for realistic values of $Q_p$ this is
still much longer than a Hubble time. 

\begin{figure}[!h]
\epsf{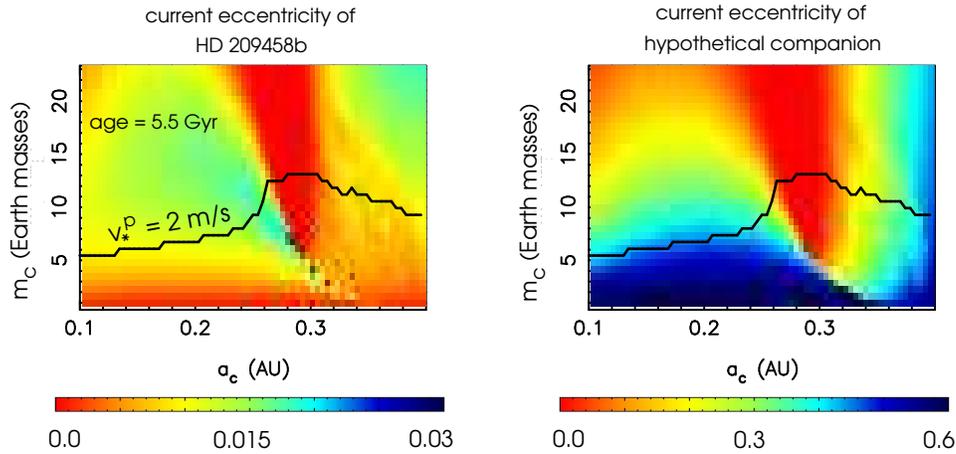}
 \caption{Eccentricities of HD 209458b and hypothetical 
 companion at 5.5 Gyr for a range of companion masses and
 semimajor axes. $v_*^p$ is the reflex velocity of the star due to the companion when it
 is at periastron. All systems except for those in the top right corners are of the type shown
 in Figure 2. Note non-zero values of HD 209458b eccentricity for $m_c<1\,M_\oplus$.
 }
 \label{conearth}
 \end{figure}

Figure~\ref{conearth}  
demonstrates that these ``pathological'' solutions are not
actually rare, existing for a significant range of values of the companion mass and semimajor axis.
Notice the extremely low-masses associated with some non-zero values of the observed planet's
current eccentricity (assuming a system age of 5.5 Gyr).
``Normal'' (exponentially decaying) systems exist for larger companion masses and semimajor axes.
Such low-mass solutions are consistent with the planet migration studies of \citet{r1}
and \citet{f1}.

\end{document}